\documentclass[prb,epsfig,superscriptaddress,twocolumn]{revtex4}
\usepackage{amsmath}
\usepackage{graphicx}

\begin{document}

\title{Stabilization mechanism of edge states in graphene}

\author{K. Sasaki}
\email{sasaken@imr.tohoku.ac.jp}
\affiliation{Institute for Materials Research, Tohoku University, 
Sendai 980-8577, Japan}

\author{S. Murakami}
\affiliation{Department of Applied Physics, University of Tokyo, Hongo,
Bunkyo-ku, Tokyo 113-8656, Japan}

\author{R. Saito} 
\affiliation{Department of Physics, Tohoku University and CREST, JST,
Sendai 980-8578, Japan}

\date{Received 29 July 2005; accepted 23 January 2006}

\begin{abstract}
 It has been known that edge states of a graphite ribbon are
 zero-energy, localized eigenstates.
 We show that next nearest-neighbor hopping process decreases the
 energy of the edge states at zigzag edge with respect to the Fermi
 energy. 
 The energy reduction of the edge states is calculated analytically 
 by first-order
 perturbation theory and numerically.
 The resultant model is consistent with the peak of recent scanning
 tunneling spectroscopy measurements.
\end{abstract}


\maketitle

Carbon-based materials have attracted much attention from various points
of view.~\cite{SDD}
In particular, their electrical properties are of great interest, where
topology plays an important role because it is relevant for a rich
variety of the electronic properties.
As one topological aspect, boundaries can induce localized states called
edge states~\cite{Fujita,Klein} at graphite edge. 
Theoretically, edge states are zero-energy eigenstates relative to the
Fermi energy and are predicted to make a certain magnetic
ordering.~\cite{Fujita,OO}
Experimentally, by scanning tunneling microscopy (STM) and spectroscopy
(STS) of graphite edge, a peak in the local density of states (LDOS) has
been observed, and it can be identified as the edge
states.~\cite{Niimi,Kobayashi,Klusek,Klusek2}
An interesting point is that the peak is located not just at the Fermi
energy but {\it below} the Fermi energy by about 20
meV.~\cite{Niimi,Kobayashi}
Since several possible perturbations would shift the energy eigenvalue
{\it above} the Fermi energy,~\cite{Klusek,Klusek2} it is not a simple
problem to find a consistent perturbation that can decrease or stabilize
the energy of the edge state.
In this letter we show that the next nearest-neighbor (NNN) hopping
process is a key factor which decreases the energy eigenvalue of the
edge state. 
This is shown by the first order perturbation theory for the
tight-binding Hamiltonian and by numerical energy band structure
calculation.

\begin{figure}[htbp]
 \begin{center}
  \includegraphics[scale=0.4]{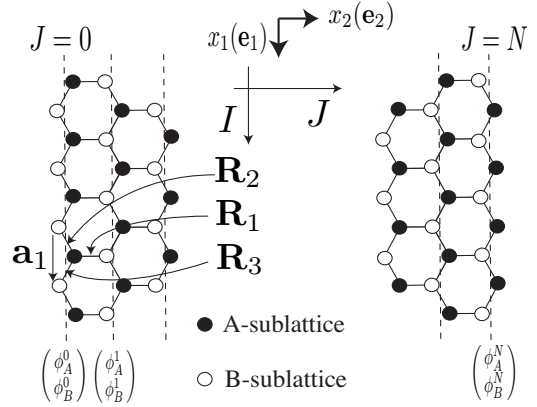}
 \end{center}
 \caption{Lattice structure of a zigzag carbon nanotube with finite
 length. 
 The filled (open) circle indicates the A sublattice (B sublattice).
 Both the left and right edges are zigzag edges.
 $\mathbf{a}_1 \equiv \sqrt{3} a_{\rm cc} \mathbf{e}_1$ is the primitive
 vector around tubule axis where $a_{\rm cc}$ denotes the
 carbon-carbon bond length and $\mathbf{e}_1$ ($\mathbf{e}_2$) is the
 unit vector around (along) the tubule axis. 
 } 
 \label{fig:edge}
\end{figure}

First, we explain the energy band structures for the graphene, using the
nearest-neighbor tight-binding Hamiltonian for the honeycomb lattice 
with the hopping integral $-\gamma_0$ ($\sim -3$ eV). 
We ignore the electron spin for simplicity. 
The eigenenergies of this model are given by 
$\pm \gamma_0 |f(\mathbf{k})|$ where 
$f(\mathbf{k}) \equiv \sum_{a=1,2,3} e^{i\mathbf{k}\cdot \mathbf{R}_a}$
and $\mathbf{R}_a$  $(a=1,2,3)$ are the vectors from an A-sublattice
site to the neighboring B-sublattice sites. 
The two energy bands are degenerate at the two $\mathbf{k}$ points
called the K and K$'$ points. 
Now we consider a zigzag nanotube illustrated in Fig.~\ref{fig:edge}.
Henceforth we define coordinate axes around and along the nanotube axis
as $x_1$ and $x_2$.
In a zigzag nanotube (of finite length) the dimensionless wave vector
around the tube, 
$q \equiv \mathbf{k} \cdot \mathbf{a}_1 =\sqrt{3} k_{1} a_{\rm cc}$,
remains a good quantum number and is now quantized as an integer
multiple of $2\pi/n$ with the chiral vector $C_h = (n,0)$.
For $n\rightarrow \infty$ the model represents a graphite ribbon with
zigzag edges discussed in Ref.~\onlinecite{Fujita}.
The delocalized eigenstates for the nanotube are similar to those of the
graphene. 
We take the Brillouin zone to be $0\le q < 2\pi$.
The states at the Fermi level without doping are characterized by 
$f(\mathbf{k})=0$, which implies $q =2\pi/3,\ 4\pi/3$, corresponding to
the K and K$'$ points, respectively.

On the other hand, as we noted in the introduction, the zigzag nanotube
has edge states in addition to delocalized (bulk) states.
Whether or not such edge states
are allowed depends on the boundary conditions. 
To analyze the edge states, we consider the eigenvalue equation
\begin{align}
 \begin{split}
  & -E \phi_A^J = \gamma_0 \phi_B^{J+1} + G \phi_B^J, \\
  & -E \phi_B^{J+1} = \gamma_0 \phi_A^J + G \phi_A^{J+1}, 
 \end{split} \ \ (J=0,\ldots,N-1)
\end{align}
where $G \equiv 2\gamma_0 \cos \left( q/2 \right)$ and 
${}^t(\phi_A^J,\phi_B^J)$ is the wave function at $J$th site
(see Fig.~\ref{fig:edge});
the wave function at site $(I,J)$ of A sublattice (B sublattice) can be
written as $\psi_{(A,B)I}^J(q) = \exp(iI q) \phi_{(A,B)}^J(q)$.

The energy eigenvalues for the nanotube with zigzag edges can be
calculated by imposing the boundary condition: 
$-E \phi_A^N =  G \phi_B^N$ and $-E \phi_B^0 =  G \phi_A^0$.
This boundary condition for the tube supports edge states around
the Fermi energy.~\cite{Fujita}
By direct calculation one finds that the edge states exist if 
$|G| < \gamma_0$, 
implying $2\pi/3<q< 4\pi/3$, and the wave function of the edge
state for $-\gamma_0 < G <0$ is
\begin{align}
 \begin{split}
 & \phi_A^J =
  \left[ \frac{\sinh (J+1) \varphi}{\sinh \varphi} \right] \phi_A^0,
  \\
 & \phi_B^J = 
  \left[ \frac{\gamma_0}{G} \frac{\sinh J \varphi}{\sinh \varphi} 
  + \frac{\sinh (J+1) \varphi}{\sinh \varphi} \right] \phi_B^0,
  \label{eq:edge-wf}
 \end{split} \ \ (J=0,\ldots,N)
\end{align}
where $\varphi$ is a positive number satisfying
$\gamma_0 \sinh(N+1) \varphi +G \sinh (N+2)\varphi = 0$ and depends on
the wave vector around the tube via the boundary condition:
$e^{-\varphi} \approx -2 \cos (q/2)$. 
The energy eigenvalue is  
$E = \pm \gamma_0 \sinh\varphi/\sinh(N+2)\varphi$,
which is exponentially close to the Fermi level.
When $\varphi$ is large, the edge state asymptotically behaves
exponentially near the edges
\begin{align}
 \begin{split}
  0 \lesssim J \ll N &:
  \phi_A^J \sim 0, \ \ \phi_B^J \sim \phi_B^0 e^{-J\varphi},
  \\
  0\ll J\lesssim N &: \phi_A^J\sim \phi_A^N e^{(J-N)\varphi}, \ \
  \phi_B^J \sim 0,
  \label{eq:asym}
 \end{split}
\end{align}
where $\phi_{B}^{0}=\mp \phi_{A}^{N}$.
The localization length is given by $\varphi^{-1} (3a_{\rm cc}/2)$.
The energy eigenvalues are 
$E \approx \pm \gamma_0 e^{-N \varphi} \approx 0$.
For $0 < G <\gamma_0$, the edge state wave function is obtained by multiplying
$(-1)^J$ by $\phi_A^J$ and $\phi_B^J$ in Eq.~(\ref{eq:edge-wf}) with
$G \to -G$. 
When approaching $G\rightarrow \pm \gamma_0$, i.e., 
$q \rightarrow 2\pi/3,4\pi/3$, 
the localization length becomes infinite,  
and the edge state finally becomes a bulk state. In this sense,
the zero-energy states at $q =2\pi/3,4\pi/3$ can be called critical
states.~\cite{SMSK}
On the other hand, the state with $q=\pi$ corresponds to $\varphi=\infty$;
it is the most localized state, which has nonzero amplitude only at the edge
sites.

Thus far we have considered the nearest-neighbor hopping process only. 
Now, we include the NNN hopping process with hopping integral
$-\gamma_n$. 
We show that the inclusion of the NNN hopping decreases the edge state
energy with respect to the Fermi level.

First, we prove this by means of the first-order perturbation theory.
The Hamiltonian matrix reads as 
\begin{align}
 {\cal H}= 
 \begin{pmatrix}
  -\gamma_n \left( |f(\mathbf{k})|^{2}-3 \right) &
  -\gamma_0 f^{*}(\mathbf{k}) \cr
  -\gamma_0 f(\mathbf{k}) & 
  -\gamma_n \left( |f(\mathbf{k})|^{2}-3 \right)
 \end{pmatrix}.
\end{align}
The new energy dispersion relation is then given by 
\begin{align}
 E(\mathbf{k}) =  \pm \gamma_0 |f(\mathbf{k})| 
 - \gamma_n |f(\mathbf{k})|^2+3\gamma_n.
 \label{eq:Ennn}
\end{align}
Henceforth we subtract the constant term $3\gamma_n$ in the energy 
$E(\mathbf{k})$. 
Thus the energy shift due to the 
NNN hopping is given by $\Delta E=-\gamma_n|f(\mathbf{k})|^2$.
The critical states have $f(\mathbf{k})=0$, which yields $\Delta E=0$.
The edge states appearing in the zigzag nanotubes with long length
correspond to $f(\mathbf{k}) \approx 0$ with complex wave vector along
the nanotube $k_{2} (\equiv \mathbf{k} \cdot \mathbf{e}_2)$.
This na{\"i}vely leads to the result that the energy shift, $\Delta E$,
of the edge states due to the NNN hopping is zero.
Nevertheless, it is not correct, 
as we will see here; 
the existence of the boundaries is crucial for the calculation of energy
shift. 
The resulting shift will turn out to be negative, thus stabilizing the
edge states.

Within the first-order perturbation theory, the energy shift of the edge
state is given by the expectation value of the NNN hopping with respect
to the edge state.
As a simple example, we consider the energy shift for the most localized
edge state ($q=\pi$).
The wave function of this edge state is given by 
$\psi_{B I}^0= \psi_{A I}^N = (-1)^I/\sqrt{2n}$
and zero otherwise.
The energy shift is negative and is given by 
$- 2\gamma_n \cos q-3\gamma_n = -\gamma_n$.
In the similar way, one can evaluate the energy shift for general edge
states as a function of $q$.
From the asymptotic behavior of the wave function given by
Eq.~(\ref{eq:asym}), we evaluate the energy shift to the first-order in
the NNN hopping as  
\begin{equation}
 \Delta E\approx \gamma_n \left(-1+G^{2}/\gamma_0^2 \right)=
  \gamma_n(2\cos q+1).
  \label{eq:DE}
\end{equation}
In particular, for the critical states, it reproduces  $\Delta E=0$ as
mentioned previously. 
Among the localized states, the negative energy shift,
$|\Delta E|=-\Delta E$, is largest for the most localized state.

\begin{figure}[htbp]
 \begin{center}
  \includegraphics[scale=0.4]{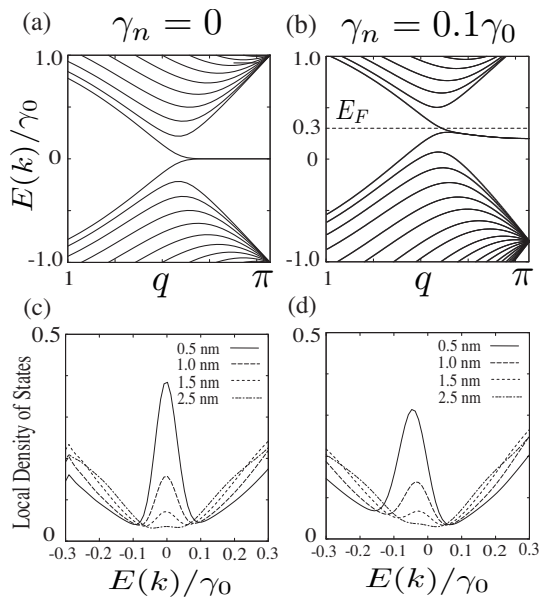}
 \end{center}
 \caption{Energy band structure of zigzag-edge, nano-graphite (a)
 without the NNN hopping and (b) with the NNN hopping 
 ($\gamma_n = 0.1 \gamma_0$). Local density of states at some points
 from the edges (0.5 - 2.5 nm) (c) without the NNN hopping and (d) with
 the NNN hopping. In (d), $E=0$ is taken as the Fermi energy.
 }
 \label{fig:band}
\end{figure}

To confirm these results, we numerically diagonalize the tight-binding
Hamiltonian for a graphite ribbon with zigzag edges.
We consider the energy band structure of a graphite ribbon with $N=20$.
In Fig.~\ref{fig:band}(a), we show the energy band structure without
the NNN hopping: $\gamma_n = 0$.
The edge states form a flatband at $E=0$~\cite{Fujita} which makes a
peak in the LDOS shown in Fig.~\ref{fig:band}(c).
In Fig.~\ref{fig:band}(b), we plot the energy band structure including
the NNN hopping.
We set $\gamma_n=0.1 \gamma_0$~\cite{Porezag} which shifts the Fermi
level to $E_F = 0.3 \gamma_0$.~\cite{fe}
The critical states are located at $E_F (=3\gamma_n)$, and the edge
states have lower energies by $-\gamma_n\le \Delta E < 0$, in agreement
with Eq.~(\ref{eq:DE}). 
Therefore, we conclude that the energy minimum at $q = \pi$ shown in
Fig.~\ref{fig:band}(b) gives a sharp peak in the LDOS.
In Fig.~\ref{fig:band}(d) we plot the LDOS at several points near the
zigzag edge as a function of energy measured from the shifted Fermi
level.
The abovementioned peak is clearly seen and this is responsible for the
peak in the LDOS observed by recent experiments.~\cite{Niimi,Kobayashi}

It is noted that our model does not include the overlapping integral
($s$ parameter~\cite{SDD}) which increases (decreases) the conduction
(valence) band width.
To examine the effect of $s$ parameter on the edge states, 
we performed 
the energy band structure calculation in an extended tight-binding
framework~\cite{Samsonidze} and found the similar behavior of the energy
band structure depicted in Fig.~\ref{fig:band} (b).
The most stable edge state is $q = \pi$ and the energy eigenvalue is
located below the Fermi level by about $\gamma_n$.
The $s$-parameter does not affect the energy spectrum near the Fermi
level since the effect of the overlapping integral is proportional to
the energy of the corresponding states measured from the Fermi level.

Here, we discuss the relationship between our work and experimental 
results~\cite{Niimi,Kobayashi,Klusek} on a peak of LDOS at graphite
edges.
Niimi {\it et al.}~\cite{Niimi} observed a clear peak at a zigzag edge
and found no such signal at an armchair edge.
The peak in the LDOS is located {\it below} the Fermi energy (defined by
zero bias voltage) by about 20 meV. 
The intensity of the peak depends on the distance from the edge, which
can be attributed to the localized nature of the edge states.
Kobayashi {\it et al.}~\cite{Kobayashi} observed a similar behavior.
In addition, they observed a peak not only at a zigzag edge but also at 
defect points of an armchair edge, while they found no peak at a 
homogeneous armchair edge.
The important point from our viewpoint of this letter is that the above
two experimental groups observed the peak located {\it below} the Fermi 
level by about $10^{-2}\gamma_0$.
However, this property does not seem to be a common property for 
edge states.
By using the STM and STS, Klusek {\it et al.}~\cite{Klusek,Klusek2}
found peaks of LDOS in the energy range of 20-250 meV {\it above} the
Fermi level at the edges of circular pits on
graphite surface. 
Although the circular pits have a mixture of zigzag and armchair edge
shapes, it is expected that there appear the edge states near (local)
zigzag edges in the circular pits since Nakada 
{\it et al.}~\cite{Nakada} showed numerically that localized states
appear not only in the zigzag edges but also in edges with other
shapes.
Such a general edge state is beyond the scope of this letter.
It is important to note that samples examined by each experimental group
were not prepared under the same condition.
Kobayashi {\it et al}.~\cite{Kobayashi} observed the graphene with clear
edge structures which were terminated with hydrogen in ultrahigh vacuum
(UHV) conditions.
Their procedure of sample preparation can exclude functional groups
including oxygen which tend to lower the energy of the edge states.
On the other hand, there is a possibility that samples used by Niimi
{\it et al}.~\cite{Niimi} and Klusek {\it et al}.~\cite{Klusek,Klusek2}
included such functional groups since their samples were not treated
with hydrogen which is activated at high temperatures in UHV conditions.

Although we have demonstrated that the NNN hopping can decrease the
energy of the edge state at zigzag edge, there is still a gap between
our result, $\Delta E = -\gamma_n \approx -10^{-1}\gamma_0$, and
experimental data,~\cite{Niimi,Kobayashi} $-10^{-2}\gamma_0$.
This gap can be attributed to several physical origins.
Among several factors, the Coulomb interaction 
would give a significant charging energy to the localized edge states.
The energy shift due to the NNN hopping depends on the localization
length of the edge state and varies from $-\gamma_n \le \Delta E < 0$. 
The most localized edge states have the largest energy 
shift  $\Delta E \approx -\gamma_n$; as the 
localization length becomes longer, the energy shift approaches
$\Delta E \approx 0$. 
Because the Coulomb charging energy is basically inversely proportional
to the localization length, the energy shift of the edge state relative
to the Fermi level will be reduced.
It is noted that the Coulomb interaction is of particular importance
from the point of view of the spin polarization.~\cite{Fujita}

In summary, we point out that the NNN hopping process
decreases the energy of edge states at zigzag edges with respect to the 
Fermi energy.
The energy reduction depends on the localization length of the edge
states. The most localized edge states 
have the largest energy reduction due to the NNN hopping, 
while the critical states stay on the Fermi level when
the sample is sufficiently large.
We calculate the energy shift [Eq.~(\ref{eq:DE})]
by the first-order perturbation theory, 
and confirm the result numerically as shown in 
Fig.~\ref{fig:band}.

K.S. acknowledges support form the 21st Century COE Program of the
International Center of Research and Education for Materials of Tohoku
University.
S.M. is supported by Grant-in-Aid (No.~16740167) from the Ministry of
Education, Culture, Sports, Science and Technology (MEXT), Japan.
R.S. acknowledges a Grant-in-Aid (No. 16076201) from MEXT.


\begin{thebibliography}{20}
\bibitem{fe}
	 Strictly speaking, the Fermi energy depends on the size of the
	 graphite and is located somewhere between $2\gamma_n$ and
	 $3\gamma_n$.
	 However, if the graphite is sufficiently large 
	 $N \gg \gamma_0/\gamma_n$ the Fermi level can be regarded as
	 $3\gamma_n$ since the number of the bulk states is much larger
	 than that of the edge states.
\end{thebibliography}
\end{document}